\documentclass[fleqn,usenatbib]{mnras}

\usepackage{newtxtext,newtxmath}
\usepackage{graphicx}
\usepackage{comment}
\usepackage{subfigure}
\usepackage{braket}

\newcommand{\fesc}{\(f_{esc}\)}

\newcommand{\afesc}{\(\langle f_{esc} \rangle \)}

\newcommand{\MUV}{\(M_{UV}\)}

\newcommand{\lya}{Ly$\alpha$}
\newcommand{\lyb}{Ly$\beta$}
\newcommand{\hi}{H\thinspace{\sc i}}
\newcommand{\hii}{H\thinspace{\sc ii}}

\newcommand{\hb}{H$\beta$}

\newcommand{\oii}{O\thinspace{\sc ii}}
\newcommand{\oiii}{O\thinspace{\sc iii}}

\usepackage[T1]{fontenc}

\DeclareRobustCommand{\VAN}[3]{#2}
\let\VANthebibliography\thebibliography
\def\thebibliography{\DeclareRobustCommand{\VAN}[3]{##3}\VANthebibliography}

\usepackage{graphicx}	
\usepackage{amsmath}	
\usepackage[]{threeparttable}
\graphicspath{{./}{figures/}}

\title[Empirical reionization model]{An Empirical reionization history model inferred from the low-redshift Lyman continuum survey and the star-forming galaxies at $z>8$}

\author[Y. Lin et al.]{
Yu-Heng Lin,$^{1,2}$\thanks{E-mail: lin00025@umn.edu}
Claudia Scarlata,$^{1,2}$
Hayley Williams,$^{2}$
Wenlei Chen,$^{2}$
Patrick Kelly,$^{1,2}$ 
Danial Langeroodi,$^{3}$
\newauthor
Jens Hjorth,$^{3}$
John Chisholm,$^{4}$
Anton M. Koekemoer,$^{5}$
Adi Zitrin,$^{6}$
and Jose M. Diego,$^{7}$
\\
$^{1}$School of Physics and Astronomy, University of Minnesota, 116 Church St SE, Minneapolis, MN 55455, USA\\
$^{2}$Minnesota Institute for Astrophysics, University of Minnesota, 116 Church St SE, Minneapolis, MN 55455, USA\\
$^{3}$DARK, Niels Bohr Institute, University of Copenhagen, Jagtvej 128, 2200 Copenhagen, Denmark\\
$^{4}$Astronomy Department, University of Texas at Austin, 2515 Speedway, Stop C1400, Austin, TX 78712-1205, USA\\
$^{5}$Space Telescope Science Institute, 3700 San Martin Dr., Baltimore, MD 21218, USA\\
$^{6}$Physics Department, Ben-Gurion University of the Negev, P.O. Box 653, Beer-Sheva 8410501, Israel\\
$^{7}$IFCA, Instituto de F\'isica de Cantabria (UC-CSIC), Av. de Los Castros s/n, 39005 Santander, Spain\\
}

\date{Accepted XXX. Received YYY; in original form ZZZ}

\pubyear{2023}

\begin{document}
\label{firstpage}
\pagerange{\pageref{firstpage}--\pageref{lastpage}}
\maketitle

\begin{abstract}
We present a new analysis of the rest-frame UV and optical spectra of a sample of three $z>8$ galaxies discovered behind the gravitational lensing cluster RX\,J2129.4+0009. We combine these observations with $z>7.5$ galaxies from the literature, for which similar measurements are available. 
As already pointed out in other studies, the high [\oiii]$\lambda$5007/[\oii]$\lambda$3727 ratios ($O_{32}$) and steep UV continuum slopes ($\beta$) are consistent with the values observed for low redshift Lyman continuum emitters, suggesting that such galaxies contribute to the ionizing budget of  the intergalactic medium. 
We construct a logistic regression model to estimate the probability of a galaxy being a Lyman continuum emitter based on the measured \MUV, $\beta$, and $O_{32}$. Using this probability and the UV luminosity function, we construct an empirical model that estimates the contribution of high redshift galaxies to reionization. The preferred scenario in our analysis shows that at $z\sim8$, the average escape fraction of the galaxy population (i.e., including both LyC emitters and non-emitters) varies with \MUV, with intermediate UV luminosity ($-19<M_{UV}<-16$) galaxies having larger escape fraction.   
Galaxies with faint UV luminosity ($-16<M_{UV}<-13.5$) contribute most of the ionizing photons. The relative contribution of faint versus bright galaxies depends on redshift, with the intermediate UV galaxies becoming more important over time.  UV bright galaxies, although more likely to be LCEs at a given log($O_{32}$) and $\beta$, contribute the least of the total ionizing photon budget.   
\end{abstract}




\section{Introduction}
Reionization occurs rapidly at redshift $z\gtrsim6$ \citep{Becker_2001, Fan_2006, Banados_2018, Eilers_2018, Planck_2018, Becker_2021}, where most of the neutral hydrogen in the intergalactic medium (IGM) was ionized by the first sources of Lyman continuum ($\lambda <$ 912\,\AA) photons. 
There is substantial evidence that star-forming galaxies are the dominant source of ionizing radiation, since the number density of quasars significantly decreases at  $z > 3$ \citep{Matsuoka_2018,Kulkarni_2019,Jiang_2022, Schindler_2023}. Some uncertainties still remain about the relative role of bright versus faint galaxies \citep{Robertson_2015, Sharma_2016,Finkelstein_2019, Naidu_2020}, and on the contribution of low luminosity AGN to the overall ionizing photon budget \citep{Giallongo_2019, Jiang_2022,Lu_2022}. 

Since the launch of {\it JWST}, the multi-object Near InfraRed Spectrograph \citep[NIRSpec, ][]{Jakobsen_2022} has begun to spectroscopically confirm and characterize galaxies during (or perhaps even prior to) the epoch of reionization \citep[EoR,][]{Pontoppidan_2022, Arellano_2022, Schaerer_2022, Trump_2022, Carnall_2023,Curti_2023, Rhoads_2023, Williams_2022, Langeroodi_2022, Mascia_2023,Tang_2023}.  The rest-frame ultraviolet (UV) and optical spectra of these galaxies are observed to have steep UV continuum slopes and large [\oiii]$\lambda$5007/[\oii]$\lambda$3727 (=O$_{32}$) emission line ratios, which are typical of hard ionizing sources and of sources with escape fraction of ionizing radiation larger than zero \citep{Izotov_2021, Flury_2022, Flury_2022_2}. 

The contribution of star-forming galaxies to the reionization of the intergalactic medium (IGM) is often parameterized as a function of three components $\dot{N}_{ion} = \text{\fesc}\,\xi_{ion}\,\rho_{UV},$ where $\dot{N}_{ion}$ is the production rate of ionizing photon emitted to the IGM, \fesc\ is the escape fraction of Lyman continuum (LyC) photons, $\xi_{ion}$ is the ionizing photon production efficiency, and $\rho_{UV}$ is the UV luminosity density measured at rest-frame 1500\,\AA\ \citep{Robertson_2022}.

The UV luminosity density is the integral of the UV luminosity function down to a given magnitude limit. The luminosity function (LF), often described with a Schechter function \citep{Schechter_1976}, can be directly measured at all redshifts. Before {\it JWST}, {\it Hubble} and ground based programs provided measurements of the redshift evolution of the LFs  
out to $z\sim 8$ \citep{Bouwens_2015, Livermore_2017,Atek_2018, Bouwens_2021}. Its shape, when observations are sufficiently deep to cover the knee of the function, does not deviate substantially from a Schechter LF, although a double power-law parameterization has been suggested for the highest redshift samples at $z\sim 6$ \citep{Bowler_2015, Khusanova_2020, Harikane_2022, Donnan_2023}. Results before the first {\it JWST} data were obtained indicated that luminous galaxies are progressively less numerous toward high-redshifts. For example, luminous galaxies ($M_{UV} < -19$) at $z\sim 8$ are 25 times less numerous compared to those at $z\sim 0$ \citep{Bouwens_2021}. The early $JWST$ results have discovered more bright galaxies at the early universe at $z>8$ than we expected \citep{Finkelstein_2022}.  A double power-law LF \citep[e.g.,][]{Harikane_2022} is proposed to include the newly discovered population of bright galaxies, since the double power-law (DP) LF are higher at the bright UV end (\MUV$<-22$), and decreases slower toward high redshift at $z>8$ than the LFs proposed pre-$JWST$ \citep[e.g.,][]{Bouwens_2021}.

$\xi_{ion}$ is the ionizing production rate relative to the UV luminosity density at 1500\,\AA. 
$\xi_{ion}$ depends on the shape of the galaxy's ionizing spectrum, which in turns depends on the initial mass function (IMF), the stellar metallicity, and the fraction of binary stars. During the EoR, galaxies are believed to have a ``top-heavy” IMF \citep{Dave_2008, van_Dokkum_2008, Sharda_2022}, lower metallicity, and low dust extinction, where all these properties lead to a higher  $\xi_{ion}$ \citep{Chisholm_2019, Atek_2022}. Indeed, the average $\xi_{ion}$ is observed to increase toward higher redshifts \citep{Matthee_2017,Shivaei_2018,Atek_2022, Matthee_2022}. 
Models of the reionization history of the universe \citep[e.g.,][]{Robertson_2013} typically assume a range of log($\xi_{ion}$) between 25.2 and 25.3, depending on the specifics of the stellar population models that are used \citep[e.g.,][]{Leitherer_1995}. This range is often referred to as the ``canonical" log($\xi_{ion}$) \citep[e.g.,][]{Shivaei_2018}. Some studies find that the average log($\xi_{ion}$) evolves with redshift \citep{Matthee_2017}. The cause of this evolution is typically ascribed to the evolution of the intrinsic properties of galaxies (e.g., the lower stellar metallicity). The mode of star-formation at any given time, however, can also be important.  This is because the definition of $\xi_{ion}$ is linked to the lifetimes of  massive stars \citep{Kennicutt_2012}. Being produced by recombination in \hii\ regions, the \hb\ luminosity traces the presence of the massive stars with short lifetimes (tens of Myrs), while the UV continuum, can also be generated from intermediate mass stars, which have longer lifetimes. Accordingly, if galaxies were, on average, characterized by shorter and more frequent bursts of star-formation at higher redshifts (for example because their dark-matter halo is still growing and feedback could have a stronger impact than in galaxies at low redshift) we would expect a different $\xi_{ion}$ distribution, even for the same physical properties such as metallicity and IMF.

In the calculation of $\dot{N}_{ion}$, \fesc\ is the term that is least constrained by observations, and the only one that cannot be measured directly for galaxies seen during the EoR. At redshifts $z \gtrsim 4$, even if ionizing photons were escaping from galaxies in a proportion similar to their low redshift  counterparts, those photons would be absorbed by the neutral hydrogen in the IGM \citep[e.g.,][]{Worseck_2014}. Therefore, during the EoR, \fesc\ can only be estimated using empirical indirect indicators calibrated from low-redshift LyC emitters \citep{Flury_2022}. 
Direct observation of the intrinsically faint LyC is challenging. Until recently, only a few dozens of galaxies at $z < 0.4$ were spectroscopically confirmed as LyC emitters (\fesc\ $>$ 0) \citep[][]{Bergvall_2006, Leitet_2011, Borthakur_2014, Leitherer_2016, Izotov_2016nat, Izotov_2016MNRAS, Izotov_2018, Flury_2022}, and even fewer LyC emitters at $2 < z < 4$ have been identified \citep{Vanzella_2016, deBarros_2016,Shapley_2016,Bian_2017, Rivera-Thorsen_2019, Fletcher_2019, Marques-Chaves_2021, Saldana-Lopez_2022_VANDELS}. Using indirect probes of LyC escape fraction, such as high values of O$_{32}$, steep UV continuum slopes, and high star formation rate surface densities, various authors have been successful in identifying a population of LyC emitting galaxies \citep{Chisholm_2018, Naidu_2020, Flury_2022, Saldana-Lopez_2022, Chisholm_2022}. These indicators, however,  may not directly provide the value of \fesc, as the escape is a complex process that depends on the neutral gas density and covering fraction \citep{Bassett_2019}. 

In this paper, we present a new analysis of a sample of gravitationally lensed galaxies at $z>8$ discovered in the RX\,J2129 galaxy-cluster field \citep{Langeroodi_2022, Williams_2022}. We refer to these three galaxies as the RXJ2129 high-$z$ galaxies.
Using these galaxies and inference based on the analysis of the low-redshift LyC survey by \citet{Flury_2022}, we present a new empirical model for the galaxy contribution to the reionization history. 
The structure of the paper is as follows. In Section~\ref{sec:data}, we describe the observations and our measurements.  We compare our objects with the low redshift Lyman continuum emitters in Section~\ref{sec:LzLCE}.  In Section~\ref{sec:Reionization}, we present our estimation of reionization history and conclude in Section~\ref{sec:conclusion}. Throughout the paper, we denote \fesc\ as the singular and plural absolute Lyman continuum ``escape fraction" and ``escape fractions", and we assume a $\Lambda$CDM cosmology with $H_0=$ 67.66 km s$^{-1}$ Mpc$^{-1}$, $\Omega_b=$ 0.04897, and $\rho_c=$8.5988$\times$10$^{-30}$ g cm$^{-3}$ \citep{Planck_2018}.

\section{Observations and Analysis} \label{sec:data}

The details of the observations and the data reduction are reported in the companion papers \citet{Williams_2022} and \citet{Langeroodi_2022}. Briefly, we obtained imaging of the RXJ2129 cluster field  with the {\it JWST} NIRCam instrument in the F115W, F150W, F200W, F277W, F356W, and F444W filters as part of the Director's Discretionary program (DD-2767; PI: P. Kelly) to obtain follow-up spectroscopy of the strongly lensed SN 2022riv \citep{kellyzitrinoguri22}.  We identified three high-redshift galaxy candidates using the EAZY \citep{Brammer_2008} photometric redshift estimation algorithm.  The follow-up spectroscopy of the RXJ2129 cluster field was obtained using the NIRSpec instrument on {\it JWST} in Multi-Object Spectroscopy (MOS) mode.   The spectrum wavelength covers from 0.6$\mu$m to 5.3$\mu$m, and the spectral resolution ranges from $R \approx$ 50 on the blue end to $R \approx$ 400 on the red end.  

The flux calibration for the NIRSpec spectroscopy was performed in the PHOTOM step of the {\it JWST} Spec2Pipeline\footnote{https://jwst-pipeline.readthedocs.io/}. Aperture corrections were applied to the NIRSpec data in the PATHLOSS step of the Spec2Pipeline. This step calculated the expected slit-losses for a point source in a given position within the shutter. Since our sources are so small with half light radius $R_e\simeq0.04\pm0.01$ arcsecond \citep{Williams_2022}, we did not apply any additional aperture correction.

\begin{figure*}
	\centering
        \includegraphics[width=1.0\textwidth]{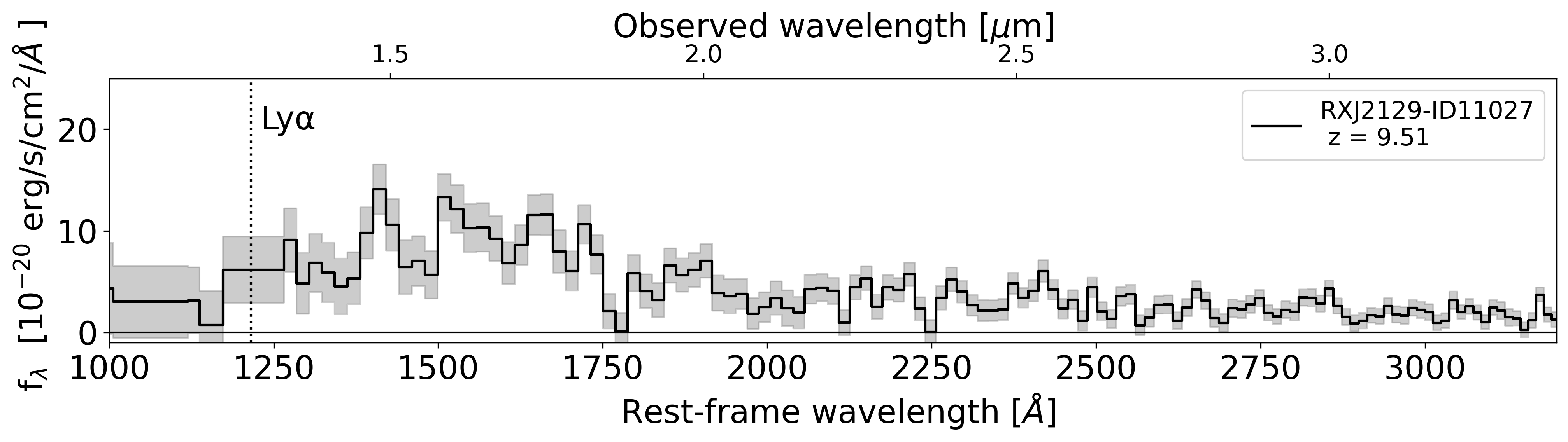}
	\includegraphics[width=1.0\textwidth]{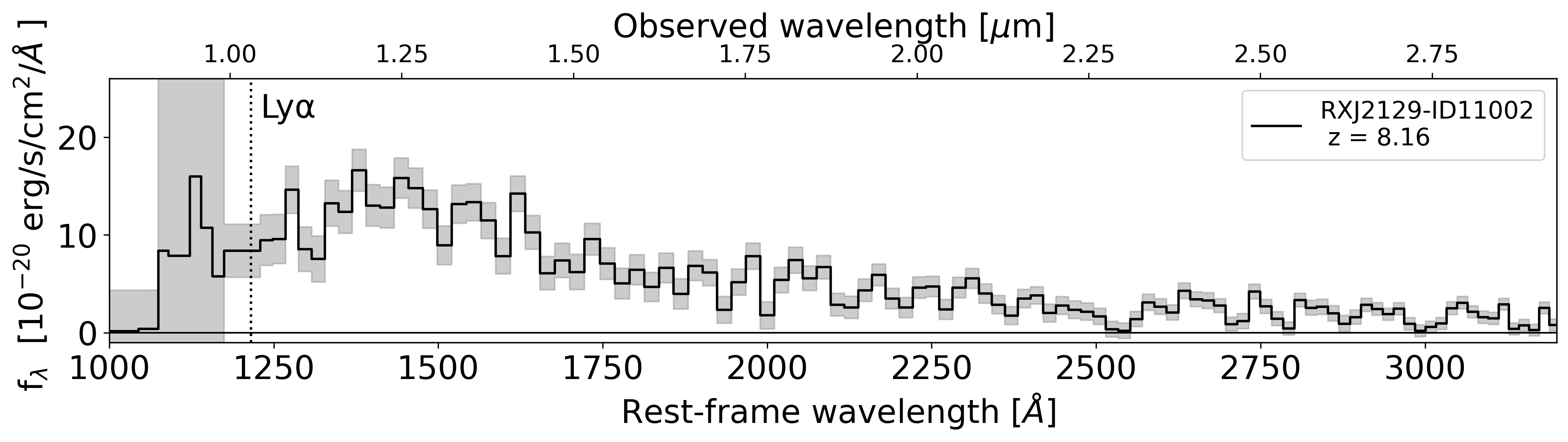}
        \caption{The rest-frame UV spectra of RXJ2129-ID 11027 at $z$=9.51 (top) and  RXJ2129-ID 11002 at $z$=8.16 (bottom).  The black lines are the spectra and the gray shaded areas are the uncertainty. 
        In the RXJ2129-ID 11002 spectrum, the masked peak at 1125\,\AA\ is  caused by a cosmic-ray hit.  }
        \label{fig:UV_spec}
\end{figure*}

In Figure~\ref{fig:UV_spec}, we show the rest-frame UV spectra with associated errors of RXJ2129-ID 11027 at redshift $z$=9.51, and  RXJ2129-ID 11002 at $z$=8.16. The rest-frame UV spectrum of RXJ2129-ID 11022 falls outside of the spectral range covered by the detector. Note that in the RXJ2129-ID 11002 spectrum, the masked peak at 1125\,\AA\ is caused by a cosmic-ray hit rather than \lya\ emission.

\subsection{Ultraviolet Properties } 
In this section we describe how we derive the galaxy properties which will be needed  in Section~\ref{sec:Reionization} to compute the average ionizing background (i.e., to compute $\dot{N}_{ion}$), namely the 1500\AA\ absolute UV magnitude ($M_{UV}$), the  slope of the spectral continuum, $\beta$, the escape fraction of ionizing radiation, \fesc, and the ionizing photon production efficiency, $\xi_{ion}$.  
Other properties, such as measurements of rest-frame optical emission lines, are presented in the companion paper on the mass-metallicity relation \citep{Langeroodi_2022}.

\subsubsection{$M_{UV}$ and the slope of the stellar continuum, $\beta$ }


To measure the $M_{UV}$ we use the rest-frame UV spectrum, when available. Specifically, we measure $M_{UV}$ by averaging the flux density between 1400\,\AA\ and 1600\,\AA. This is a spectral region typically free from strong emission lines. For galaxy ID 11022,  we calculate the $M_{UV}$ using the flux density computed in the F150W filter, which corresponds to a rest-frame wavelength of 1650\,\AA. 
The analysis of the spectral energy distributions  presented in \citet{Williams_2022} and \citet{Langeroodi_2022} suggests that they suffer low level of dust attenuation. To calculate the observed  UV continuum slopes we do not apply any dust correction to the observed magnitudes. 
The color excess E(B$-$V) is then estimated with the observed UV continuum slopes \citep{Chisholm_2022}.

Since the spectrum of of galaxy ID 11027 suffers from the contamination (Figure~\ref{fig:UV_spec}), we measure the UV continuum slopes, $\beta$, defined as \(f_\lambda \propto \lambda^{\beta}\), by fitting a power-law function to the observed flux densities in the photometric bands F1500W, F200W, and F277W. We used Markov Chain Monte Carlo sampling to sample the posterior on the parameter $\beta$. 

The observed UV continuum slopes $\beta\simeq-2$ suggest a certain amount of dust attenuation level, with E(B$-$V)$_{SMC}$ $\simeq$ 0.03 \citep[see ][Section 4 and Appendix A]{Chisholm_2022}.  We adopt the Small Magellanic Cloud-like dust attenuation law from \citet{Gordon_2016} with R$_V$ $\equiv$ A$_V$/E(B$-$V) = 2.74 to calculate the dust extinction.  


\subsubsection{ Ionizing Photon Production Efficiency $\xi_{ion}$ }
The ionizing photon production efficiency $\xi_{ion}$ is defined as the ratio between the ionizing photon production rate $Q_0$ and the non-ionizing UV luminosity density at 1500\,\AA, $L_{\nu}(1500)$, $\xi_{ion} = Q_0/L_{\nu}(1500) $. 
Assuming Case~B recombination theory \citep{Osterbrock_book, Leitherer_1995}, we can write $Q_0$ in terms of the observed \hb\ luminosity, L(\hb), and the escape fraction of ionizing radiation, \fesc\ (i.e., $Q_0 \propto L(H\beta)/(1-f_{esc})$). 
Then, $\xi_{ion}$  can be estimated as:

\begin{equation}
\xi_{ion} = \frac{L(\text{\hb})}{4.76\times10^{-13} (1-f_{esc})L_{\nu}(1500)}. \label{eq:xi}
\end{equation}

\noindent
There are two sources of uncertainty in this estimation of $\xi_{ion}$: \fesc\ and the dust extinction. 
The dust extinction ($A_V \simeq 0.1$) decreases the log($\xi_{ion}$) by $\sim$ 0.13 dex.  
In order to quantify the uncertainties introduced by the unknown \fesc, 
we introduce the photon production efficiency computed for \fesc$=0$, $\xi_{ion}^0=\xi_{ion} (\text{\fesc} = 0)$.

\subsubsection{Escape Fraction $f_{esc}$ } 

The escape fraction of ionizing radiation cannot be directly measured at $z\gtrsim 4$ because the optical depth for ionizing photons is high 
\citep[e.g.,][]{Worseck_2014}. Accordingly, we need to use indirect estimators of \fesc, calibrated at lower redshifts using galaxies as close as possible to those that we observe at $z\gtrsim 6$, during the EoR.  \cite{Schaerer_2022} demonstrated that local extreme emission line galaxies have similar properties to the galaxies {\it JWST} has been uncovering. 

A number of studies have measured \fesc\ in the local universe and found relations between \fesc\ and galaxy properties.  Despite a number of \fesc\ indicators (e.g., O32, $\beta$) have successfully been identified using low-redshift LyC emitters, the scatter between the value of  \fesc\ and the value of its indirect estimator is observed to be substantial \citep{Flury_2022_2}. These indicators, therefore,  may not be sufficient for an accurate estimate of \fesc\ due the complicated escape process involving gas density and covering fraction \citep{Bassett_2019}.   Here, we use the relation between \fesc\ 
 and  $\beta_{1500}$, the UV slope between 1300\AA\, to 1800 \AA,  discussed in \citet{Chisholm_2022}.  
The relation is given as follows: 
\begin{equation}
f_{esc}(\beta_{1500}) = (1.3 \pm 0.6 )\times 10^{-4}~ 10^{(-1.22 \pm 0.10)\beta_{1500}}.
\label{eq:fesc_beta}
\end{equation}

\bigskip

We report the UV properties of the RXJ2129 high-$z$ galaxies in Table~\ref{tab:info}.
For the three galaxies in the SMACS0723 sample, we adopt the magnification, UV absolute magnitudes, and photometric UV continuum slopes reported in \citet{Schaerer_2022}. Since the \hb\ flux is not reported in \citet{Schaerer_2022}, we adopt the metallicity, emission line fluxes, and ratios from \citet{Curti_2023}. 
For completeness, we report  the properties of the SMACS0723 high-$z$ galaxies in Table~\ref{tab:info2}. 
\begin{table}
\centering 
\small
\begin{tabular}{cccc} 
\hline\hline
 RXJ2129 &    11027    &    11002  &   11022    \\
\hline
redshift $z$    &   9.51    &   8.16    &   8.15    \\
magnification $\mu$ &   19.2$\pm$3.6 &    2.23$\pm$0.15 &   3.29$\pm$0.33   \\
12+log(O/H)$^{(a)}$ &  7.47$\pm$0.10  &    7.65$\pm$0.09   &    $<$7.72   \\
O$_{32}$    &    13.28$\pm$3.75   &   13.51$\pm$5.08     &    $>$6.16   \\
\hb\ flux$^{(b)}$ &   2.17$\pm$0.16   &  0.88$\pm$0.12  &     0.16$\pm$0.08        \\
$M_{UV}$$^{(c)}$ &  $-$20.78$\pm$0.26   &   $-$20.69$\pm$0.17    &     $-$18.74$\pm$0.18  \\
$M_{UV}$$^{(d)}$ &   $-$18.20$\pm$0.28   &  $-$20.31$\pm$0.19   &     $-$17.91$\pm$0.20        \\
$\beta$  &   $-$1.92$\pm0.19$     &   $-$2.29$\pm$0.21    &  $-$2.12$\pm$0.44    \\
E(B$-$V)$_{SMC}$  &    0.04   &   0.02    &  0.03   \\
log($\xi_{ion}^0$)  &    25.60$\pm$0.11  &   25.18$\pm$0.09   &   25.28$\pm$0.17   \\
f$_{esc}$($\beta$)    &  0.06$\pm$0.03    &   0.08$\pm$0.04    &  0.05$\pm$0.02      \\
\hline
\end{tabular}
\caption{The derived properties for the RXJ2129 high-$z$ galaxies. 
$^{(a)}$ Measured with strong line calibration \citep{Izotov_2019}, the measurements of RXJ2129 ID 11022 are reported in 1-$\sigma$ limit \citep{Langeroodi_2022}.  
$^{(b)}$ Observed, in unit of 10$^{-18}$ erg s$^{-1}$ cm$^{-2}$ \citep{Williams_2022, Langeroodi_2022}. 
$^{(c)}$ Observed, without correction for magnification. For RXJ2129 ID 11022, we derive the UV magnitude from the photometry at rest frame wavelength $\sim$1650\,\AA.
$^{(d)}$ Corrected for lensing and dust extinction. 
}  
\label{tab:info} 
\end{table}

\begin{table}
\centering 
\begin{tabular}{cccc} 
\hline\hline
 SMACS       &    04590    &    06355  &    10612    \\
\hline
redshift $z$    &   8.495    &   7.664    &   7.660   \\
magnification $\mu$ &   7.9 &    1.7 &   1.7   \\
12+log(O/H) &  6.99$\pm$0.11  &    8.24$\pm$0.07   &    7.73$\pm$0.12    \\
O$_{32}$    &    12.8$\pm$1.4   &   7.4$\pm$0.3     &    23.7$\pm$6.4    \\
\hb\ flux$^{(a)}$ &   1.54$\pm$0.06   &  2.11$\pm$0.05   &     1.19$\pm$0.04        \\
$M_{UV}$$^{(b)}$ &   $-$20.29   &  $-$21.09   &     $-$20.38        \\
$M_{UV}$$^{(c)}$ &   $-$18.06   &  $-$20.51   &     $-$19.81        \\
$\beta$  &   $-$2.20$\pm$0.15     &   $-$1.96$\pm$0.22    &  $-$2.31$\pm$0.11    \\
f$_{esc}$($\beta$)    &  0.06$\pm$0.03    &   0.03$\pm$0.01    &  0.08$\pm$0.04      \\
log($\xi_{ion}^0$)   &    25.72$\pm$0.04   &   25.44$\pm$0.04    &   25.47$\pm$0.04    \\
\hline
\end{tabular}

\caption{The derived properties for the SMACS0723 high-$z$ galaxies.  
$^{(a)}$ Observed, in unit of 10$^{-18}$ erg s$^{-1}$ cm$^{-2}$ \citep{Curti_2023}. 
$^{(b)}$ Observed, including lensing \citep{Schaerer_2022}.
$^{(c)}$ Corrected for lensing \citep{Schaerer_2022}, we adopted a 0.2 magnitude uncertainty throughout the calculation. 
}  
\label{tab:info2} 
\end{table}

\section{Comparison with Low Redshift Analogs } \label{sec:LzLCE}

We compare the properties of the $z>7.5$ galaxies, including the RXJ2129 high-$z$ galaxies, the SMACS0723 high-$z$ galaxies, and the galaxies recently reported in the {\it GLASS-JWST} program \citep{Mascia_2023} and the {\it CEERS} survey \citep{Tang_2023} to those of the low-redshift galaxies studied in the Low-$z$ Lyman Continuum survey \citep[LzLCs, at $z\sim 0.3$][]{Flury_2022}. 
The LzLCs includes 66 newly observed low redshift ($z<$ 0.4) galaxies and 23 galaxies ($z<$ 0.46) in the literature \citep{Izotov_2016nat,Izotov_2016MNRAS, Izotov_2018,Izotov_2018b,Izotov_2021, Wang_2019}, 
for which the LyC emitters (LCE) are defined as galaxies with Lyman Continuum detected with 97.725$\%$ confidence. We refer to these 89 galaxies as the LzLCs sample.  In this sample, 50 galaxies are confirmed as LyC emitters (LCE), and 39 galaxies are not detected in the LyC (non-LCE).
The LzLC objects were selected to span a broad range in physical properties associated with a large probability of high escape fraction of ionizing radiation. Specifically, galaxies were selected to have high O$_{32}$ ratio, steep UV continuum slope, and high star formation rate surface density, $\Sigma_{SFR}$. 

\subsection{The probability of a galaxy being a LyC emitter}\label{sec:logistic}
\begin{figure}
\includegraphics[width=0.99\linewidth]{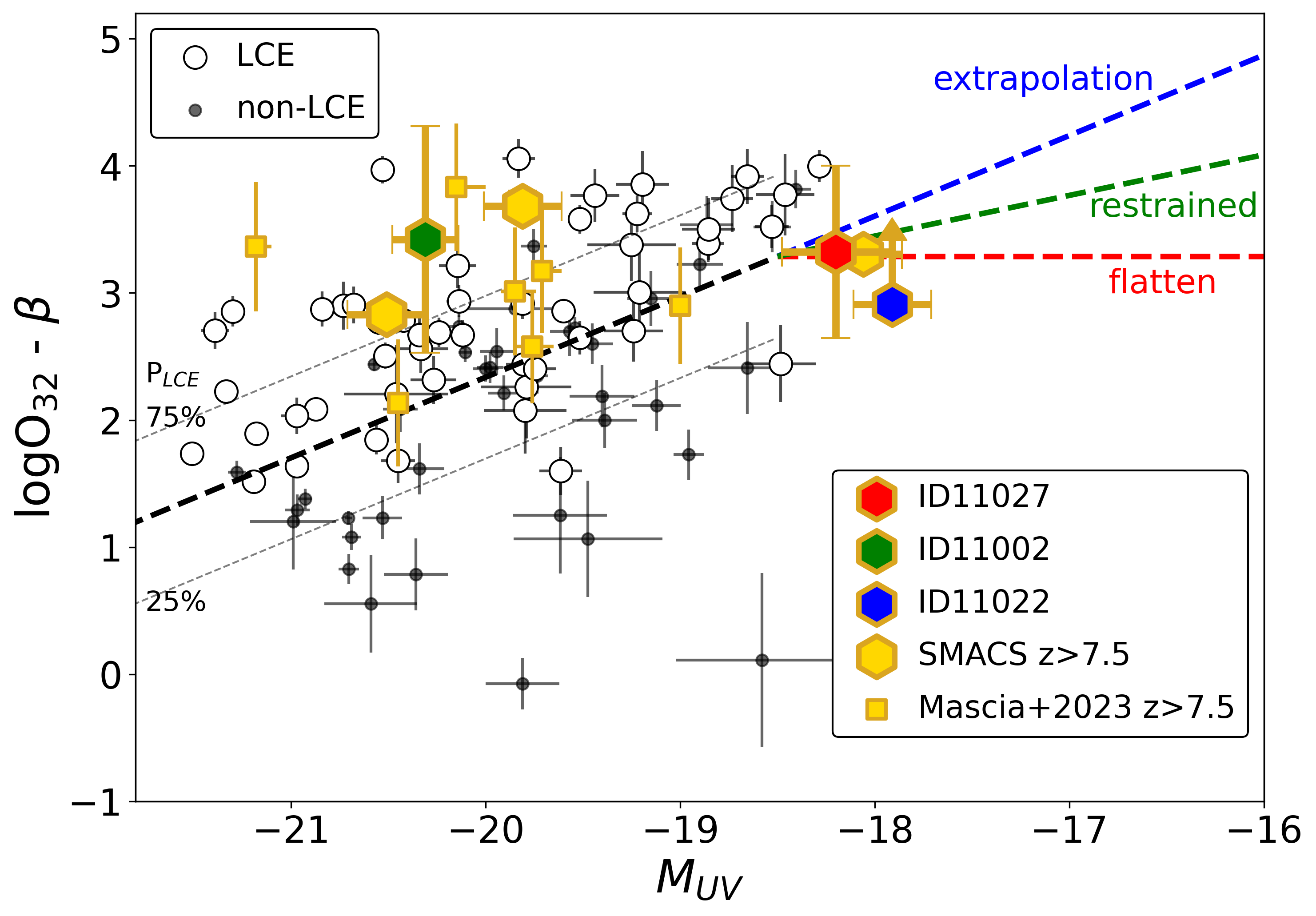}
\caption{ The combined \fesc\ indicator log(O$_{32}$)$-\beta$ as a function of the absolute UV magnitude. 
The red, blue, and green hexagons are the RXJ2129 high-$z$ galaxies presented in this work. The yellow hexagons are the SMACS high-$z$ galaxies reported in the {\it JWST} Early Release Observation \citep{Pontoppidan_2022}, and the yellow squares are the $z > 7.5$ galaxies in the {\it GLASS-JWST} program \citep{Mascia_2023}.
The low redshift LyC emitters (LCE, white circles) and non-LyC emitters (non-LCE, black dots) are separated in this Figure. 
 The black dashed line is the logistic discriminator described as equation~\ref{eq:logistic}, and the gray dashed lines indicate where the P$_{LCE}$=75$\%$ and P$_{LCE}$=$25\%$. It is unclear how this discriminator will extend at \MUV\  faint than $-18.5$.   We assume three simple extrapolations to account for this uncertainty: extrapolation (blue dashed line), restrained (green dashed line), and flatten (red dashed line). 
 \label{fig:fesc}}
\end{figure}

We combine the O$_{32}$ ratio and the slope of the UV continuum in a combined indicator, $\beta O_{32} =$log(O$_{32}$)$-\beta$.  We choose $\beta$, O$_{32}$, and \MUV\ in our model for their simple accessibility in the observations.  In Figure~\ref{fig:fesc}, we show how, in a diagram of $\beta O_{32}$ as a function of the absolute UV magnitude,  LyC emitters (LCE, open circles) are efficiently separated from  non-LyC emitters (non-LCE, filled black circles). We also explored whether a correlation exists between $\beta O_{32}$ and \fesc,  but we do not observe any simple relationship between these quantities, strengthening the conclusion that indirect estimators are mostly useful to identify LyC emitting galaxies, rather than estimating the value of \fesc.  
 
Accordingly,  we apply a logistic regression to the LzLCs sample, to estimate the probability that a galaxy is a LCE based on $M_{UV}$ and $\beta O_{32}$. The logistic discriminator can be written as: 

\begin{equation}
\text{$P_{LCE}$} =  \frac{1}{1 + e^{-(b_0 + b_1 M_{UV} + b_2 \beta O_{32})}},  \label{eq:logistic}
\end{equation}

\noindent
where $P_{LCE}$ is the probability of the galaxy being a LyC emitter, and the best fit values for ($b_0$, $b_1$, $b_2$) are ($-$25.82, $-$1.09, 1.72). 
 The discriminator is shown in Figure~\ref{fig:fesc} with the black dashed line, in the range where no extrapolation is necessary. 
Assuming the discriminator has no evolution on redshift, we find that RXJ2129 ID 11002, SMACS 06355, and SMACS 10612 have P$_{LCE} >$  0.8, suggesting that these galaxies are likely LyC emitters. SMACS 04590 has higher O$_{32}$ and steeper $\beta_{1500}$ than SMACS 06355, but its position in Figure~\ref{fig:fesc} (most right yellow point at $M_{UV}$ $\approx -$18) suggests that SMACS 04590 is less likely to be a LCE due to the fainter UV magnitude, with $P_{LCE}$ = 0.39. 
We note, however, that the local LzLC sample does not extend to magnitudes fainter than $\approx -18.5$,  such that it is not clear where is the boundary between LyC and non-LyC emitters at the faint end of the UV magnitude $>-19$, and our logistic discriminator may not be applicable at the faint UV side. 
The physical connections between  the \fesc\ indicators and \MUV\ is likely the changes of metallicity, dusts, and star formation.  The faint star-forming galaxies (associated with the low metallicity galaxies) have noticeably different properties compared with their bright counterpart.  For instance, the intrinsic UV continuum slope does not get appreciably bluer below 10$\%$ solar metallicity \citep{Bouwens_2010,Topping_2022}. The changes in O$_{32}$ are also less obvious below 10$\%$ solar metallicity \citep{Curti_2017, Sanders_2021}.   Therefore the $\beta$ and $O_{32}$ may become less effective as indicators for LyC leaking at the faint UV end. 
In this paper, we apply the logistic discriminator to \MUV<-18, where the fit is not constrained. To account for the uncertain behavior at the faint end we consider three scenarios, as illustrated in Figure~\ref{fig:fesc}: The "extrapolation scenario", represented by the blue dashed line, involves a simple linear extrapolation of the regression to faint magnitudes. In this scenario, faint galaxies are less likely to be classified as Lyman Continuum Emitters (LCEs), as being an LCE requires a large value for $log(O_{32}) - \beta$.  The "restrained scenario", denoted by the green dashed line, exhibits a shallower slope compared to the extrapolation scenario. In this case, the criteria for classifying faint galaxies as LCEs are less strict.  Finally, the discriminator plateaus in the "flatten scenario", indicated by the red dashed line, where faint galaxies have a significantly higher chance of being categorized as LCEs.  The parameters ($b_0$, $b_1$, $b_2$) in equation~\ref{eq:logistic} for the restrained scenario are ($-$15.83, $-$0.55, 1.72), while for the flatten scenario, the values are ($-$5.655, 0, 1.72).


The choice of the discriminator's turning point is primarily determined by observations extending down to \MUV$=-18.5$.  The slope of the extended discriminator fundamentally governs the contribution of the faint galaxy population to reionization. 
We will demonstrate in Section~\ref{sec:model_result} that observational constraints are more in agreement with the reionization history  derived from the restrained scenario.


\subsection{The dependency of $\xi_{ion}$ on $M_{UV}$ and redshift}
\begin{figure}
\includegraphics[width=0.99\linewidth]{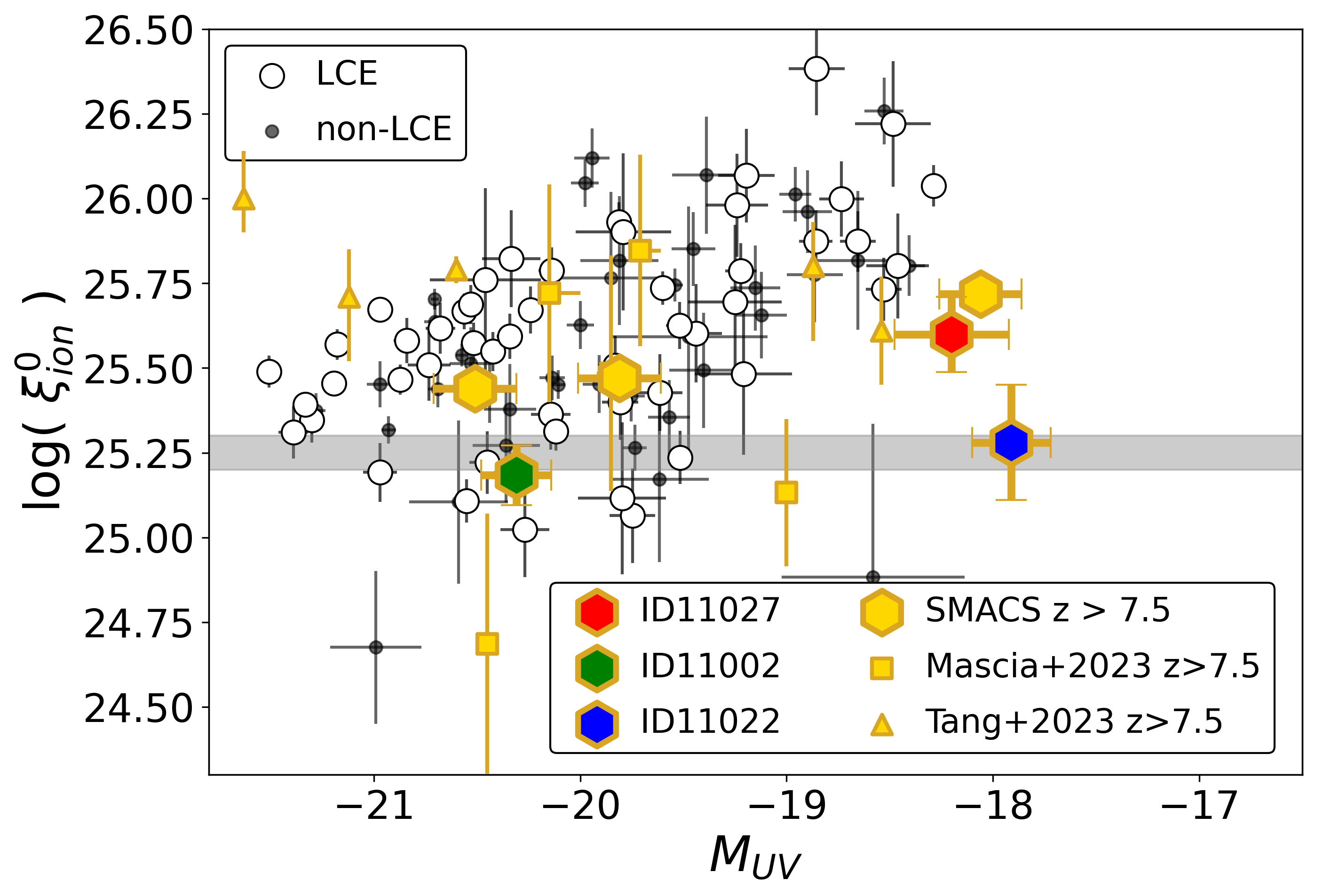}
\caption{ Same as Figure~\ref{fig:fesc} but $\xi_{ion}^0$ as a function of the absolute UV magnitude, where $\xi_{ion}^0$ = $\xi_{ion}$(\fesc=0). The gray shaded area marks the ``canonical" log($\xi_{ion}^0$) = 25.2 $--$ 25.3.  The yellow squares are the $z > 7.5$ galaxies in the {\it GLASS-JWST} program \citep{Mascia_2023}, and the yellow triangles are the $z > 7.5$ galaxies in the {\it CEERS} survey program \citep{Tang_2023}
 \label{fig:xi_vs_UV}}
\end{figure}

\begin{figure}
\includegraphics[width=0.99\linewidth]{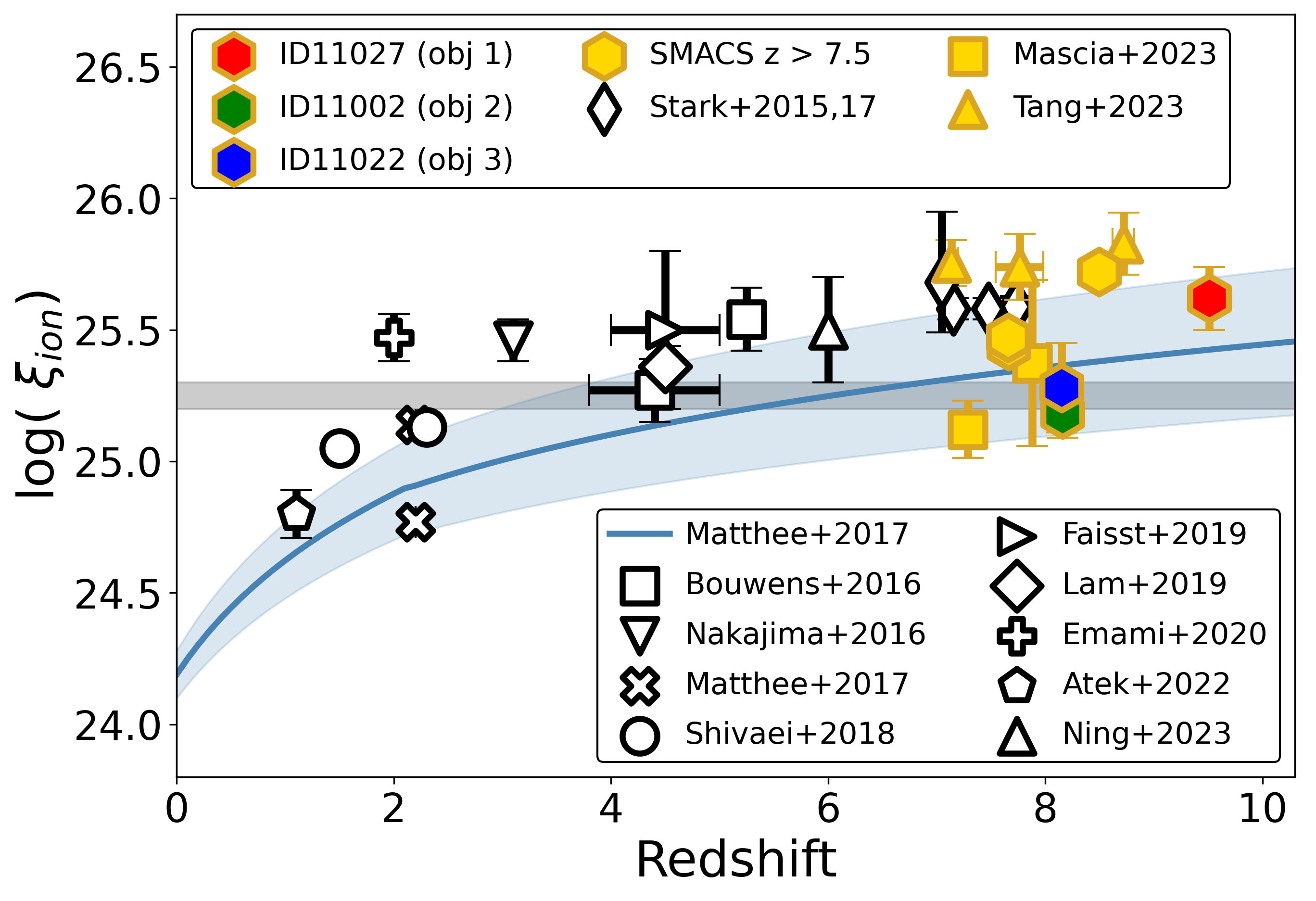}
\caption{ Ionizing photon production efficiency measured over a wide range of redshift.  The blue line and blue shaded area is the redshift evolution derived in \citet{Matthee_2017}, and the 1-$\sigma$ limit.  
 \label{fig:xi_redshift}}
\end{figure}

Many works have studied the dependence of $\xi_{ion}$, and whether $\xi_{ion}$ depends on the UV magnitudes is still under debate \citep{Matthee_2017, Shivaei_2018,Emami_2020,Nakajima_2020, Atek_2022}. 
We calculate the $\xi_{ion}^0$ of the LzLCs sample using equation~\ref{eq:xi}.
We calculate the $\xi_{ion}^0$ of the $z > 7.5$ galaxies in the {\it GLASS-JWST} program \citep{Mascia_2023} using equation~\ref{eq:xi} with \hb\ luminosities derived from the star formation rate density.
We show the log($\xi_{ion}^0$) as a function of the UV magnitude in the lower panel of Figure~\ref{fig:xi_vs_UV}, where we observe a positive correlation between $\xi_{ion}^0$ and the UV magnitude for both the LzLCs sample and the high-redshift galaxies.  However, this $\xi_{ion}^0$ vs.\ UV correlation may be a secondary dependence and the $\xi_{ion}^0$ is in fact dependent on other factors such as metallicity and burstiness. 
In Figure~\ref{fig:xi_redshift}, we show all high-redshift galaxies have log($\xi_{ion}^0$) $\geq$ 25.2, consistent with the extrapolation at $z = 8$ of the observed redshift evolution \citep{Stark_2015, Bouwens_2016, Nakajima_2016, Stark_2017, Matthee_2017, Shivaei_2018, Faisst_2019,Lam_2019, Atek_2022,Ning_2023}.

\section{ Reionization History }\label{sec:Reionization}

In this section, we use the insights  gained in the previous sections to build a new model of the reionization history. Specifically, in the calculation of the redshift evolution of the neutral fraction ($x_{\text{\hi}}$) we consider (1)  the empirical constraints on the probability that a galaxy is a LCE, (2) the dependency of \fesc\  on $\beta$, and (3) the dependency of log($\xi_{ion}$) on  $M_{UV}$ or redshift $z$.

\subsection{The empirical model}
We calculate the neutral fraction $x_{\text{\hi}}$ by solving 
\begin{equation} \label{eq:reionization}
\frac{d (1 - x_{\text{\hi}) }}{dt} = \frac{\dot{N}_{ion}}{n_{\text{H}}} - \frac{( 1- x_{\text{\hi}})}{t_{\text{rec}}}, 
\end{equation}
where $\dot{N}_{ion}$ is the ionizing photon production rate, $n_{\text{H}}$ is the comoving gas number density, and $t_{\text{rec}}$ is the recombination time scale \citep{Madau_1999,Robertson_2013, Ishigaki_2018}.  $n_{\text{H}}$ and $t_{\text{rec}}$ are defined as 
\begin{equation}
n_{\text{H}} = \frac{ X_p\Omega_b\rho_c }{m_{\text{H}}}, 
\end{equation}
\begin{equation}
t_{\text{rec}} = [ C_{\text{\hii}}~\alpha_B(T) (1 + Y_p/4X_p) n_{\text{H}} (1+z)^3  ]^{-1}, 
\end{equation}
where $X_p=0.76$, $Y_p=0.24$ are the primordial mass fraction of hydrogen and helium \citep{Planck_2018}, $\Omega_b$ is the baryon energy density fraction, $\rho_c$ is the critical density, $C_{\text{\hii}} \equiv$ $\langle n_{\text{\hii}^2} \rangle$/$\langle n_{\text{\hii}}\rangle^2$ is the clumping factor, and $\alpha_B(T)$ is the case B recombination coefficient. Here we assume $C_{\text{\hii}} = 3$, and $\alpha_B=2.6\times10^{-13}$ cm$^3$s$^{-1}$, for an electron temperature of $10^4$K. We solve equation~\ref{eq:reionization} iteratively, assuming the boundary condition that $x_{\text{\hi}} = 1$ at $z=15$, i.e., we assume that the first sources of ionizing photons appear at this redshift.\footnote{We note that  starting at $z=15$ or $z=20$ introduces a negligible difference in our model.} 

The ionizing photon production rate depends on the escape fraction of ionizing radiation, the ionizing photon production efficiency and the volume density of UV luminosity ($\dot{N} =f_{esc}\,\xi_{ion}\,\rho_{UV}$). In what follows we explain how we include the empirical constraints derived in the previous sections in the calculation of $\dot{N}$ at each time step.

\begin{figure*}
\centering
\includegraphics[width=0.99\linewidth]{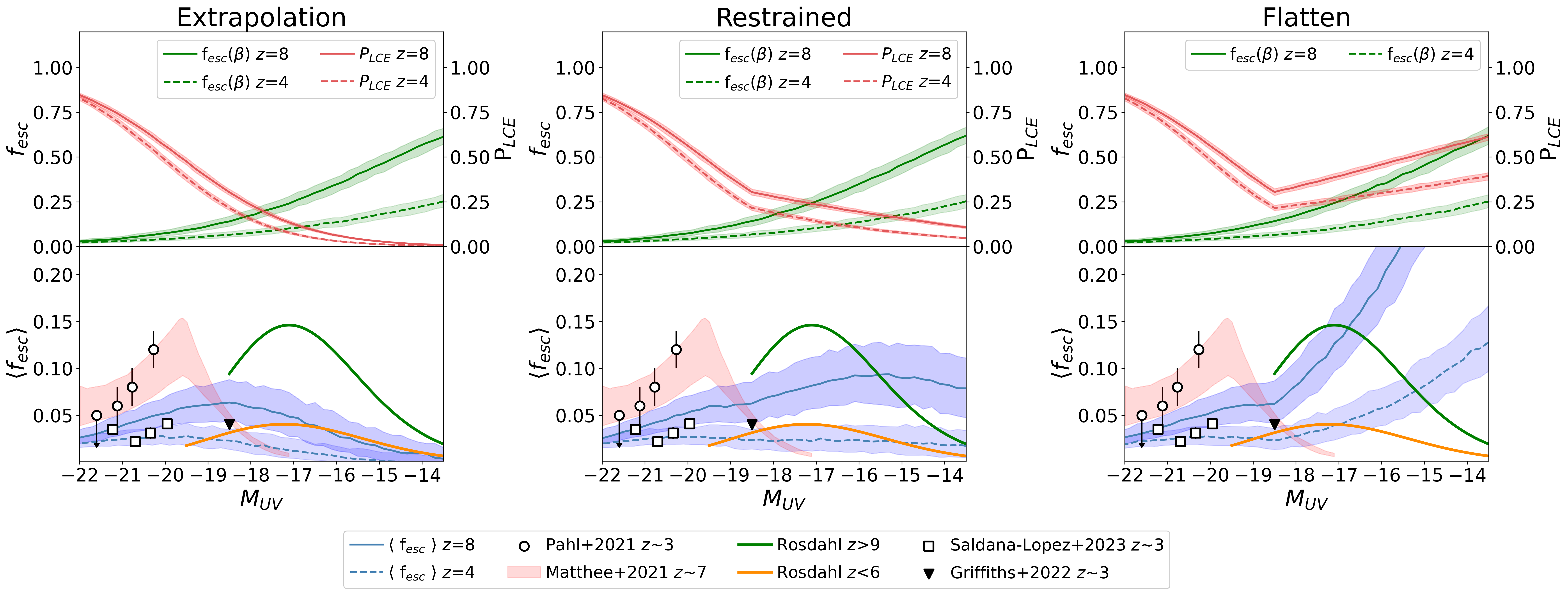}
\caption{ $P_{LCE}$ and \fesc($\beta$) as a function of absolute UV magnitude for extrapolation scenario, the restrained scenario, and the flatten scenario. Top panel: $P_{LCE}$ (red) and \fesc($\beta$) as a function of absolute UV magnitude. $P_{LCE}$ is the probability of whether a galaxy is a LCE or not. \fesc($\beta$) is the average \fesc\ value computed without accounting for $P_{LCE}$. Bottom panel: The blue line shows \afesc\, the average of \fesc. The shadow areas are the the 90$\%$ and 10 $\%$ percentiles of the 500 runs. The observational results of \citet{Pahl_2021} and \citet{Saldana-Lopez_2022_VANDELS} are displayed with white circles and squares, respectively.  The median \fesc\ from SPHINX galaxies from \citet{Rosdahl_2022} at $z > 10$ and $z < 9$ are indicated through green and orange lines, respectively.  
\label{fig:fesc_MUV} }
\end{figure*}

First, we derive the population average escape fraction \afesc\ as a function of the  UV absolute magnitude, accounting for the dependency of \fesc\ on $\beta$ and on the probability of each galaxy being a LCE. In each bin\footnote{We consider 45 bins of 0.2 magnitude between $-13.5$ and $-23$} of \MUV, we simulate 100 galaxies, each with a different value of $\beta$ and  log(O$_{32}$), drawn from the distributions described below. These values are  used, together with \MUV, to compute the probability P$_{LCE}$($M_{UV}$, $\beta O_{32}$), using Equation~\ref{eq:logistic}.  We then draw a Boolean value based on P$_{LCE}$ to determine whether or not the galaxy is a LCE. If a galaxy is a non-LCE, we set its \fesc$^i=0$. If a galaxy is a LCE, we draw its \fesc\ value from the multivariate normal distribution with parameters described by Equation~\ref{eq:fesc_beta}. 
The $\beta$ values are drawn from  observations of $z > 8$ galaxies, assuming the following normal distribution: $\beta =  ( -0.17 )\times$$M_{UV}$  $-5.4 \pm 0.4$, truncated at $\beta = -3.5$ \citep{Cullen_2022}. We maintain a fixed slope, $d\beta/dM_{UV}$, in this relation and sample $\beta$ assuming an intrinsic scatter of 0.4 dex \citep{Rogers_2014,Cullen_2022}. Properly accounting for the scatter in $\beta$ is crucial in our model  because a larger scatter results in galaxies with lower $\beta$ values, which, on average, increase both the (P$_{LCE}$) and the escape fraction.
\citet{Paalvast_2018} reported no significant correlation between stellar mass and star formation rate with the O$_{32}$ values, therefore, we assume that log(O$_{32}$) values are independent of \MUV.  The log(O$_{32}$) values are generated from a normal distribution of $0.5\pm0.1$ \citep{Sanders_2023}.   
Finally, we compute \afesc=$\frac{1}{100}\sum{}^{100}_{i=1}f^i_{esc}$ as the average \fesc\ of the 100 galaxies at each $M_{UV}$ bin, and we repeat this calculation 500 times. 
The results of our simulations for the three scenarios are presented in Figure~\ref{fig:fesc_MUV}. In the top panel, we display in red the average P$_{LCE}$ and in green the average \fesc\ computed without accounting for P$_{LCE}$, both as functions of \MUV.
The average \fesc\ without accounting for P$_{LCE}$ is identical in all three different scenarios since it depends solely on the sampled $\beta$ values generated from the same distribution. 

All three scenarios share the same discriminator for the brighter end (\MUV$<-18.5$). Therefore, the results for P$_{LCE}$ and \afesc\ for \MUV$<-18.5$ are consistent: brighter galaxies are more likely to be LCEs than fainter ones. However, since brighter galaxies have redder UV continua (larger $\beta$ values), the \fesc\ values are generally lower. These trends explain the results in the bottom panel, where we depict the average escape fraction of the population that now includes P$_{LCE}$ as a function of \MUV. For brighter galaxies (\MUV $< -18.5$), \afesc\ increases with decreasing \MUV.

For fainter galaxies, both P$_{LCE}$ and \afesc\ change in response to the discriminator's slope.  In the extrapolation scenario, most galaxies are less likely to be classified as LCEs. Consequently, the \afesc\ peaks at \MUV$=-18.5$ and decreases as \MUV becomes fainter. This \afesc\ vs.\ $M_{UV}$ pattern in our model behaves similarly to the \afesc\ inferred from the Ly$\alpha$ emitters in \citet{Matthee_2022_LAE}.   In the restrained scenario, the slope of the discriminator is lower, resulting in a larger population of faint galaxies being classified as LCEs.  In this case, the \afesc\ continues to increase until \MUV$=-16$, and the \MUV\ evolution is less pronounced compared to the previous scenario.  In the flatten scenario, the P$_{LCE}$ reverses at the faint UV end, allowing a significant fraction of faint galaxies being LCEs. As a result, the \afesc\ keeps increasing with decreasing UV magnitude.

To demonstrate the redshift evolution of \fesc, we also model the $P_{LCE}$, \fesc, and \afesc\ of the galaxies at redshift $z=4$ (dashed lines).  The $\beta$ vs.\ $M_{UV}$ dependency is adopted from \citet{Bouwens_2014} \citep[see also][]{Chisholm_2022}, and includes the intrinsic scatter $\Delta\beta\sim$ 0.4, used to draw samples of  $\beta$ for individual galaxies.  In our model, as we assume no evolution on the logistic discriminators, the $P_{LCE}$ does not evolve much with redshift. The \fesc\ and \afesc\ decrease at  lower redshifts as the UV continuum slope $\beta$ becomes flatter/redder. 
Our predicted value of \afesc$\simeq0.03$ at $z=4$ aligns with the average of \fesc=0.03$\pm$0.02 in \citet{Saldana-Lopez_2022_VANDELS}, where the \fesc\ is derived using the depth of absorbtion lines from the low-ionization metals and the UV attenuation. Our result is lower than the average of \fesc=0.06$\pm$0.01 observed with the Lyman Break galaxies at $z\sim3$ in \citet{Pahl_2021} and \fesc=0.07$\pm$0.02 at $z\sim3.5$ in \citet{Begley_2022}. The \fesc\ in \citet{Pahl_2021} are also higher than the values calculated with equation~\ref{eq:fesc_beta} using the photometric UV slope $\beta$ of the sample (see \citet{Saldana-Lopez_2022_VANDELS} Figure 15).

The \fesc\ difference can be caused by the choice of dust attenuation law, the intrinsic $\beta$ value in the stellar population model, or a potential redshift evolution \citep[see ][ for more discussion]{Saldana-Lopez_2022_VANDELS}.  
Another possible explanation of discrepancy with some of the $z\sim3$ observations is that we may have underestimated the intrinsic scatter on the $\beta$ distribution. A larger intrinsic scatter of the $\beta$ distribution would result in a greater number of individual galaxies with higher \fesc, leading to a higher \afesc.

We note that \citet{Sanders_2023} found on average log($O_{32}$)=0.9$\pm$0.1 at $6.5<z<9.3$ significantly higher than log($O_{32}$)=0.5$\pm$0.1 at $z\sim5.6$.  It is unclear whether such high log($O_{32}$) are commonly seen at $z>6$, while the $z\sim5.6$ sample has similar log($O_{32}$) to the $2.7<z<5$ galaxies. Therefore we assume the log($O_{32}$) is independent with redshift and choose to sample log($O_{32}$) with the mean of 0.5.  The value of log($O_{32}$) does not affect individual \fesc\ in our model, and only slightly changes the fraction of LCEs population at \MUV$<-18$.  As a result, higher log($O_{32}$) values do not significantly increase the \afesc\ at \MUV$<-18$. 

In both the extrapolation and restrained scenarios, the \afesc\ peaks up at intermediate UV magnitudes ($-18.5<$\MUV$<-16$) and decreases toward the faint UV magnitude end. \citet{Grazian_2017,Griffiths_2022} have established an upper limit \fesc$<0.04$ at \MUV$<-20$\footnote{we convert the relative escape fraction estimates to absolute escape fraction estimates assuming E(B-V)=0.1 and the \citet{Calzetti_2000} attenuation curve, as done in \citet{Mevtric_2021, Begley_2022}}. Although there are no sufficient observations of faint galaxies yet to study this trend, some simulations have predicted a similar pattern,  with \fesc\ peaking at intermediate UV magnitudes, rather than at the brightest or faintest end. For example, the medium \fesc\ of SPHINX galaxies in \citet{Rosdahl_2022} has the highest \fesc\ at \MUV$\sim-16$, while \citet{Ma_2020} predict that the highest \fesc\ should be in 10$^8$ stellar mass galaxies, or \MUV$\sim-19$ \citep{Stefanon_2021}.  In these studies, the decline in \fesc\ for larger and brighter objects is attributed to dust attenuation, while the decreasing for lower mass and fainter objects is attributed to inefficient  star formation and increased susceptibility to stellar feedback.



For log($\xi_{ion}$), we propose two models: the $\xi_{ion}$(UV) model and the $\xi_{ion}$($z$) model.  
In the $\xi_{ion}$(UV) model, we assume $\xi_{ion}^0$ is linearly dependent on $M_{UV}$.
The linear relation is obtained with the minimum $\chi^2$ fitting of all galaxies at $z > 7.5$ in Figure~\ref{fig:xi_vs_UV}:  $\xi_{ion}^0$(UV) = 0.11($M_{UV} + 20.0) + 25.46$, and flatten at $\xi_{ion}^0$ = 24.5 and 26. We exclude the brightest object in our fitting, as it may potentially be powered by an AGN \citep{Mainali_2018, Tang_2023}.
In the second model, we adopt $\xi_{ion}(z)$ as a function of redshift. We use the 1-$\sigma$ upper limit redshift dependence in \citet{Matthee_2017}: log($\xi_{ion}$)($z$) = 24.493+1.180 log(1+$z$), which is consistent with more $z>6$ objects in Figure~\ref{fig:xi_redshift}. 

To explore the impact of the newly discovered bright galaxy populations at $z>8$ to reionization, we adopt two luminosity functions at $z>8$ into our models: 
a Schechter LF \citep{Bouwens_2021}, and a Double Power-law (DP) LF \citep{Harikane_2022}. We find the IGM neutral fractions at $z>8$ derived the DP LF are only 1$\%$ lower than that of the Schechter LF (see Figure~\ref{fig:history}). 
Therefore, throughout this paper, we adopt the Schechter luminosity function described in \citet{Bouwens_2021}, with 
\begin{equation}
\alpha  = -1.94 - 0.11(z - 6),
\end{equation}
\begin{equation}
\Psi = 0.40\times 10^{-3}~10^{(-0.33(z-6)+(-0.024(z - 6)^2))},
\end{equation}
\begin{equation}
M_\ast = -21.03-0.04(z - 6).
\end{equation}
We integrate the UV magnitude from $-23$ to $-13.5$. The UV magnitude is truncated at $-13.5$ to match previous studies \citep{Livermore_2017, Ishigaki_2018, Atek_2018, Naidu_2020, Trebitsch_2022}. 
Using the UV mass-light ratio: log(M$_*$) = -0.49(\MUV+20.5) +8.8 at $z=8$ in \citet{Stefanon_2021}, we  estimate a stellar mass of log(M$_*$) $\simeq 5.4 M_\odot$ at \MUV$=-13.5$.

\subsection{The results of the empirical model}\label{sec:model_result}

\begin{figure*}
    \centering
    \subfigure[]{\includegraphics[width=0.48\textwidth]{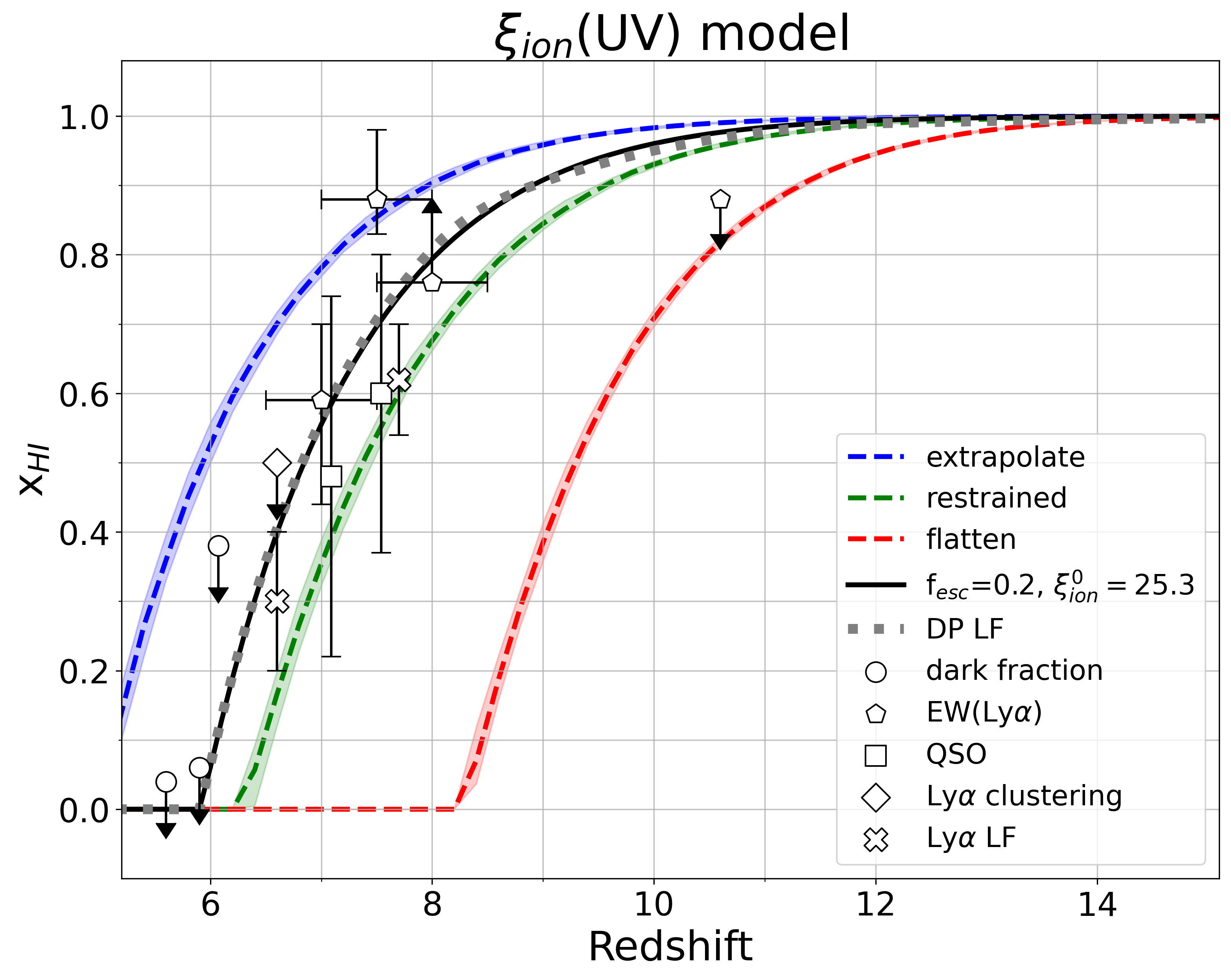}}\label{fig:history_UV} 
    \subfigure[]{\includegraphics[width=0.48\textwidth]{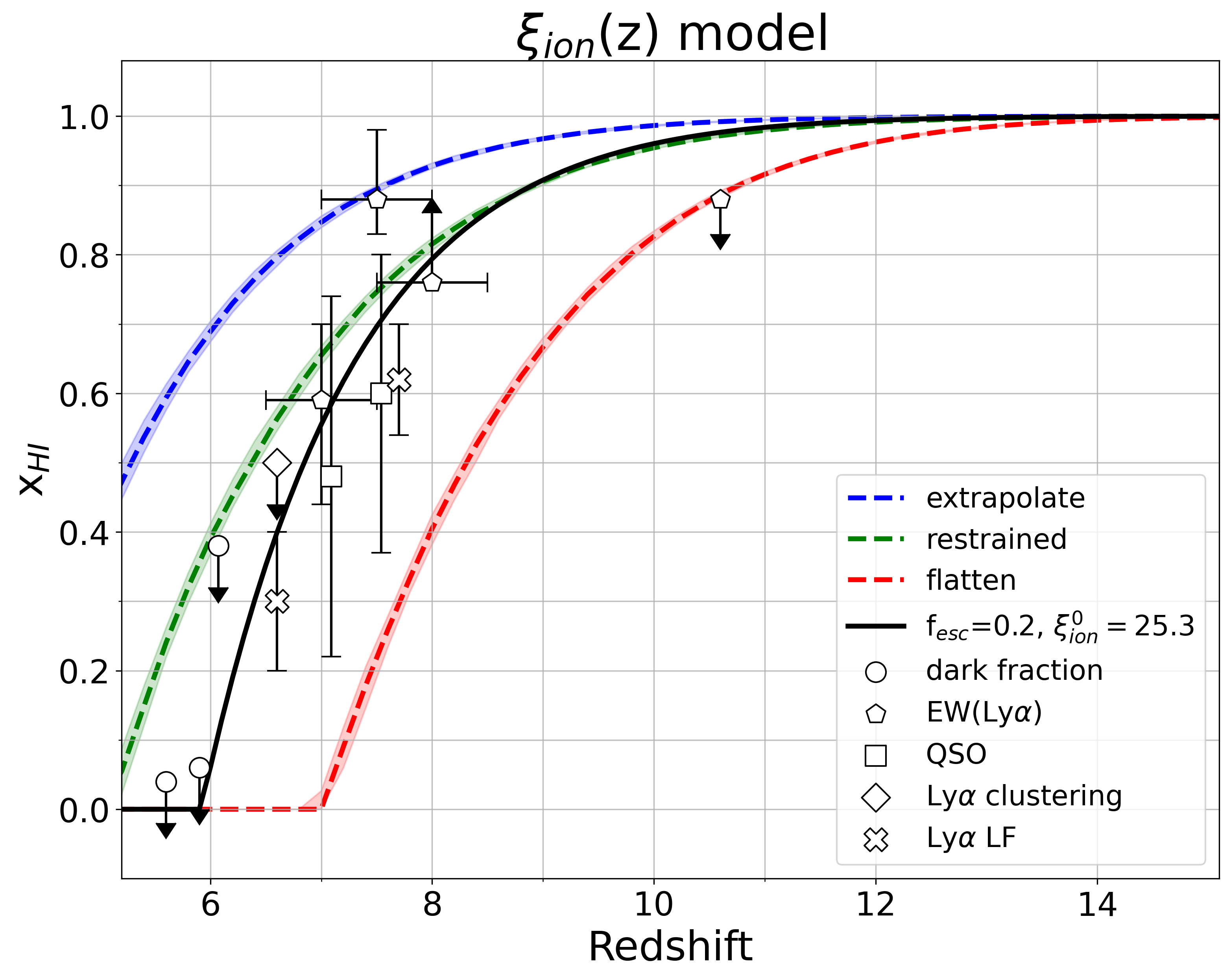}}\label{fig:history_z} 
    \caption{ Reionization History of (a) the $\xi_{ion}$(UV) model and (b) the $\xi_{ion}$(z) model. The blue, green, and red dashed lines are results based on the linear extrapolation, restrained, and flatten discriminator, respectively. The black line and red dotted line are simple model with constant \fesc = 0.2, constant $\xi_{ion}^0$ = 25.3, and single Schechter LF from \citet{Bouwens_2021} and double power-law LF from \citet{Harikane_2022}, respectively.   The shaded areas are the 90$\%$ and 10 $\%$ percentiles of the 500 runs. The observational constraints are shown in white markers.  }
    \label{fig:history}
\end{figure*}

We present the reionization history of the three scenarios (blue: extrapolation; green: restrained; red: flatten) in two models ($\xi_{ion}$(UV) and $\xi_{ion}$($z$) in Figure~\ref{fig:history_UV}, with the constraints derived from observations: 
\lya\ equivalent width (EW(\lya)) of galaxies \citep{Mason_2018,Mason_2019, Hoag_2019, Bruton_2023}; 
the clustering of \lya\ emitter galaxies \citep{Ouchi_2010,Greig_2017}; 
\lya\ and \lyb\ dark fraction \citep{McGreer_2015}; 
\lya\ luminosity function \citep{Faisst_2014,Ning_2022};
QSO damping wings \citep{Davies_2018}. 
As reference, we show two simple models where all galaxies have uniform constant \fesc = 0.2 and $\xi_{ion}^0$ = 25.3 ($\xi_{ion}$ = 25.4), adopting either Schechter luminosity function or a double power-law luminosity function.

In both models, the restrained scenario (green dashed line)  provides the closest match with observations, although the $\xi_{ion}$($z$) shows a later reionization history.  For our models, we define the beginning redshift of the reionization as $z_{90}$ when the IGM neutral fraction $x_{\text{\hi}}$ = 0.9. For the $\xi_{ion}$(UV) model, $z_{90}\sim 9.6$, and for the $\xi_{ion}$($z$) model $z_{90}\sim 9.0$.  We define the ending redshift of the reionization as $z_{10}$ with $x_{\text{\hi}}$ = 0.1. For the $\xi_{ion}$(UV) model, $z_{10}\sim 6.4$, and for the $\xi_{ion}$($z$) model $z_{10}\sim 5.4$.  

\begin{figure*}
    \centering
    \subfigure[]{\includegraphics[width=0.48\textwidth]{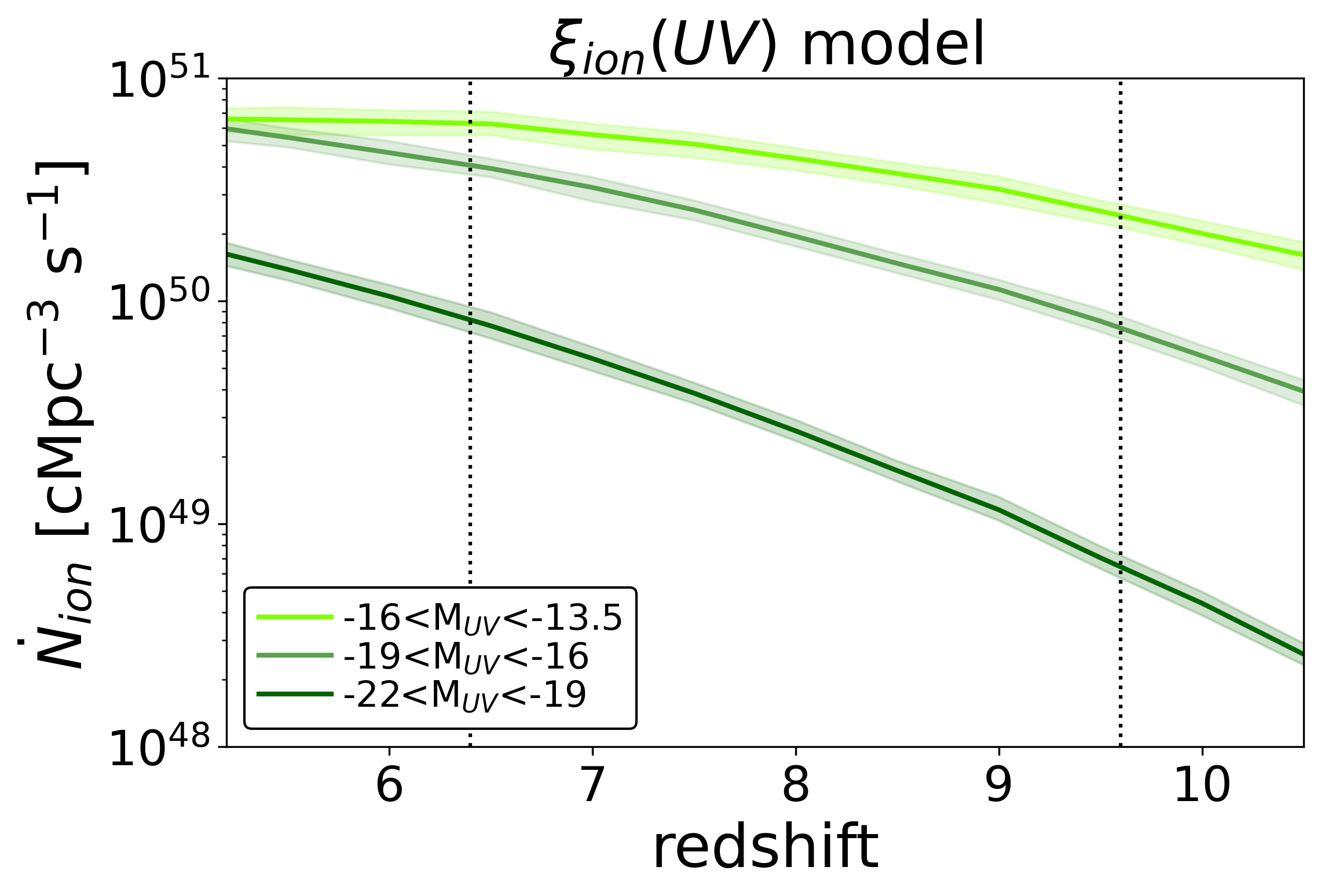}}
    \subfigure[]{\includegraphics[width=0.48\textwidth]{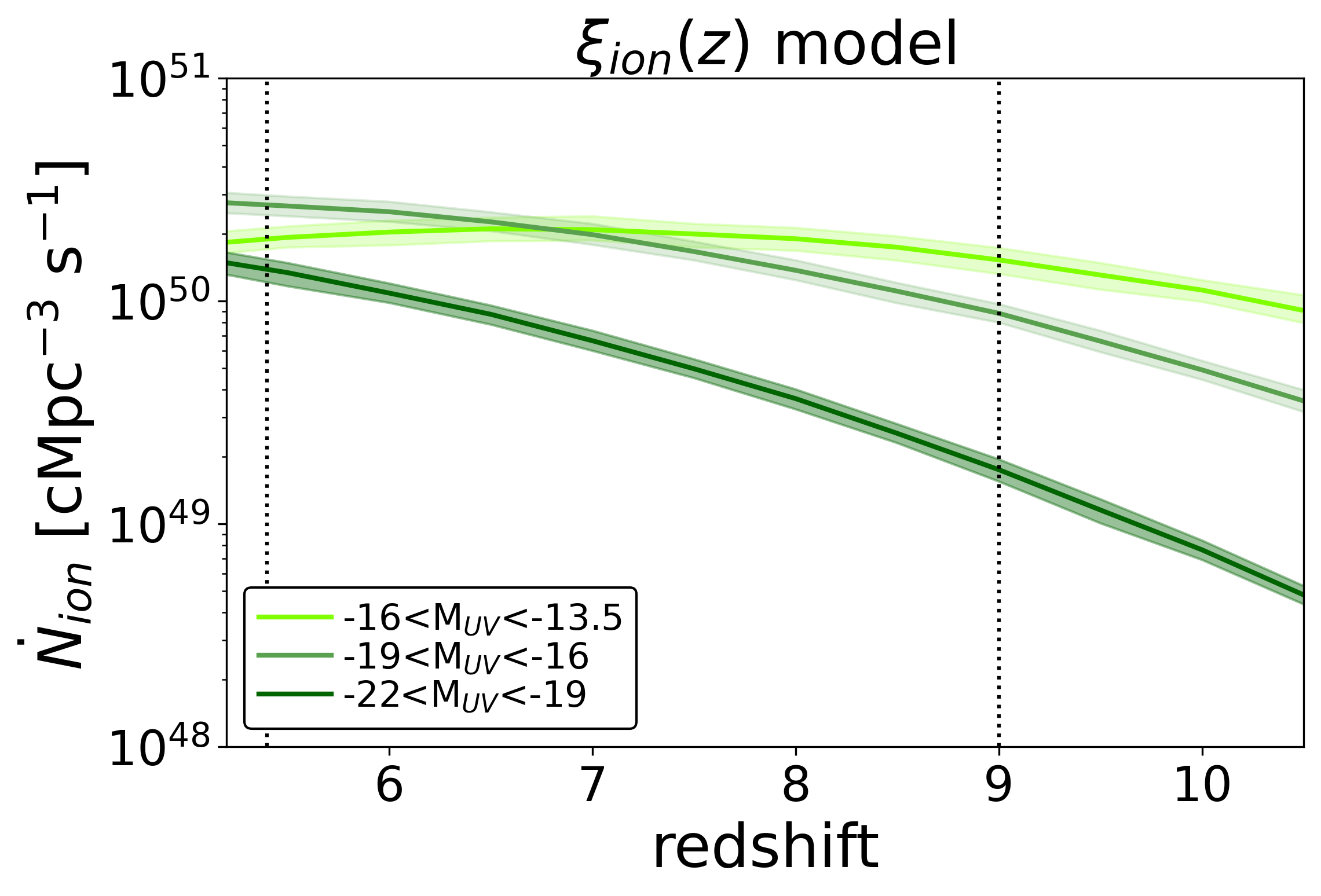}}
    \caption{ The $\dot{N}_{ion}$ of galaxies grouped by $M_{UV}$ as a function of redshift in the restrained scenario.  The light to heavy green lines show the contribution from the UV faint ($-16<$$M_{UV}<-13.5$), intermediate ($-19<$$M_{UV}<-16$), and bright ($-23<$$M_{UV}<-19$) galaxies, respectively.   The shaded areas show the 90$\%$ and 10 $\%$ percentiles of the 500 runs.  The vertical dashed lines are the redshifts $z_{90}$ and $z_{10}$, when the IGM neutral fractions $x_{\text{\hi}}$ are 0.9 and 0.1, respectively.    }
    \label{fig:contribution}
\end{figure*}

In Figure~\ref{fig:contribution} we focus on the restrained scenario and compute, for both models, the contribution to the reionization from galaxies with different UV luminosities. We group  galaxies by their $M_{UV}$ into UV faint ($-16<M_{UV}<-13.5$), UV intermediate ($-19<M_{UV}<-16$), and UV bright ($-23<M_{UV}<-19$) galaxies. In both models, the UV faint galaxies are the major contributors of the reionization, producing $\sim 50\%$ of the ionizing photons at all redshifts. The contribution of the UV intermediate galaxies increases as time goes by, becoming comparable to that of the UV faint galaxies at end of the reionization. UV bright galaxies, even though more likely to be LCEs at a given log($O_{32}$) and $\beta$, contribute only 10$\%$ to 20$\%$ of the total ionizing photon budget.  

The distribution of log($O_{32}$) and $\beta$ can affect the reionization history and the relative contribution of galaxies with different UV luminosity.  If the log($O_{32}$) has a higher mean, as reported in \citet{Sanders_2023}, more UV faint galaxies will be classified as LCEs in the restrained and the flatten scenarios, but not in the extrapolation scenario. The relative contribution of UV faint galaxies will be even higher in this case. If the scatter of the $\beta$ is larger, individual UV bright galaxies with very blue $\beta$ can bring up \afesc\ in all scenarios, while the UV intermediate and faint galaxies do not change as much since the $\beta$ value is truncated at $-$3.5. In this case the relative contribution of UV bright galaxies in all scenarios will be higher.

\section{Conclusions}\label{sec:conclusion}
In this paper we present a new analysis of the rest frame UV and optical spectra of a new a sample of $z >$~8 galaxies discovered behind the gravitational lensing cluster RX\,J2129.4+0009 \citep{Williams_2022, Langeroodi_2022}. We combine these observations with those of the $z>7.5$ galaxies for which similar data are available  \citep{Pontoppidan_2022, Arellano_2022, Schaerer_2022, Trump_2022, Carnall_2023,Curti_2023, Rhoads_2023, Mascia_2023, Tang_2023}.

We compare the properties of these galaxies with those observed as part of the low redshift Lyman continuum survey \citep{Flury_2022}. The high [\oiii]$\lambda$5007/[\oii]$\lambda$3727 emission line ratios (O$_{32}$) and steep UV continuum slopes, $\beta < -2$, of our sample are consistent with the values observed for low redshift Lyman continuum emitters, suggesting that these galaxies potentially contribute to the ionizing budget for  the intergalactic medium. We use the \hb\ and UV luminosity to estimate the average ionizing photon production efficiency of our sample.

We apply a logistic regression (equation~\ref{eq:logistic}) to estimate the probability of a galaxy with \MUV$<-18$ being a Lyman continuum emitter based on the measured \MUV\ and $\beta O_{32}$ values, and explore three scenarios to account for the uncertain behavior at the faint end. Using this probability, we construct an empirical model that estimates the galaxy contribution to the reionization budget based on the observable quantities (\MUV, $\beta$, $O_{32}$).  
The preferred scenario in our analysis shows that at $z=8$, the average escape fraction of the galaxy population (i.e., including both LyC emitters and non-emitters) varies with \MUV. 
\fesc\ is approximately 4\% for bright galaxies ($M_{UV}$ $< -$19), and peaked at intermediate UV luminosity ($-19<M_{UV}<-16$). Galaxies with faint UV luminosity ($-16<M_{UV}<-13.5$) contribute half of the ionizing photons throughout the epoch of reionization. The relative contribution of faint versus bright galaxies depends on redshift, with the intermediate UV  galaxies becoming more important over time. UV bright galaxies, even though more likely to be LCEs at a given log($O_{32}$) and $\beta$, contribute the least of the total ionizing photon budget.   
 

\section*{Acknowledgements}
P.L.K. is supported by NSF grant AST-1908823 and anticipated funding from {\it JWST} DD-2767. D.L. and J.H. were supported by a VILLUM FONDEN Investigator grant (project number 16599). A.Z. acknowledges support by Grant No.\ 2020750 from the United States-Israel Binational Science Foundation (BSF) and Grant No. 2109066 from the United States National Science Foundation (NSF), and by the Ministry of Science \& Technology, Israel.

\section*{Data Availability}

The data used in this article will be shared on reasonable request to the corresponding author.



\bibliographystyle{mnras}
\bibliography{manuscript} 








\bsp	
\label{lastpage}
\end{document}